# Modelling Automotive Function Nets with Views for Features, Variants, and Modes

H. Grönniger[2], J. Hartmann[1], H. Krahn[2], S. Kriebel[1], L. Rothhardt[1], B. Rumpe[2]

1: BMW Group, München, Germany
2: Institute for Software Systems Engineering, TU Braunschweig, Germany

**Abstract**: Modelling the logical architecture of an automotive system as one central step in the development process leads to an early understanding of the fundamental functional properties of the system under design. This supports developers in making design decisions. However, due to the large size and complexity of the system and hence the logical architecture, a good notation, method and tooling is necessary. In this paper, we show how logical architectures can be modelled succinctly as function nets using a SysML-based notation. The usefulness for developers is increased by comprehensible views on the complete model that describe automotive features, variants, and modes.

**Keywords**: Automotive Logical Architecture, Variants, Modes, SysML

## 1. Introduction

Developing automotive embedded systems is a complex task since a large number of different functions from several vehicle domains interact in many ways. As for software functions which are the main driver for innovations in the automotive domain, experts estimate that, by 2010, one gigabyte of onboard-software will be integrated in premium class vehicles [8].

Mature methodologies and processes are needed to break down the complexity into manageable tasks. Likewise important is a good tool support to efficiently perform the development tasks like requirements engineering for vehicle features, planning software and hardware architecture or designing controllers and implementing the software.

In this paper, we assume a development process as depicted in Figure 1. Different teams that are responsible for developing a certain feature (i.e., functionality perceptible by customers) capture the requirements mostly textually with the help of a requirements management system (RMS), e.g. DOORS [14]. A software and hardware architecture, a mapping between these two, and the final realisation are derived from this abstract form of specification in subsequent working stages. These are ideally developed by applying the AUTOSAR methodology [1] and related tools.

As an important step in between, we assume the modelling of a logical architecture since the frequently found direct transition from feature requirements to hardware and software architecture raises some problems. Specifying software typically involves the decision which logical functions can be integrated in a single software function or have to be split to more than one software function. Design decisions about which logical function is implemented in hardware or software have to be made. Without an explicit logical architecture, a consistent functional integration becomes difficult and high coordination efforts are needed, especially if several suppliers are involved. The main reason is the complexity of automotive software architecture which makes a more abstract representation necessary to document the main functionalities and design decisions in a readable comprehensible model.

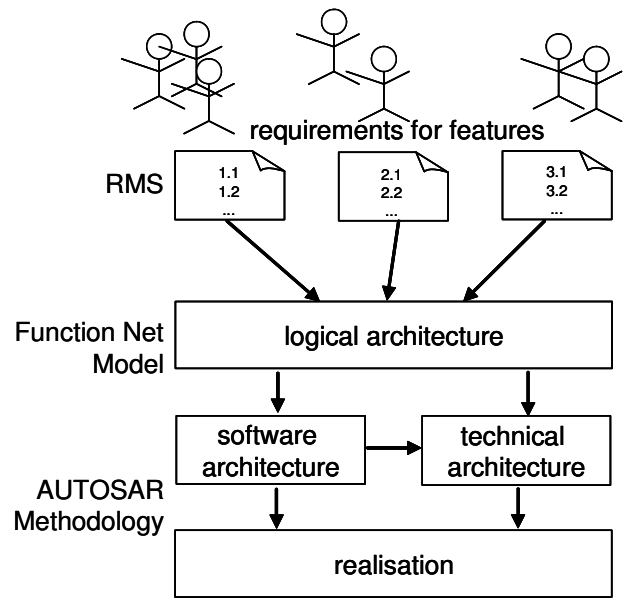

**Figure 1** Development Process

To be practically useful, logical architectures have to fulfil the following requirements.

1. The logical architecture provides a system description that is more abstract than descriptions on the software and hardware architecture level. It should not contain technical details.

2. Automotive systems are typically not developed from scratch. Development documents from previous product cycles form the basis for following development activities. Being able to re-use and to adapt logical architectures hence is a key requirement for the desired productivity and quality gain of using logical architectures.
3. That also means that logical architectures must provide a comprehensible documentation of functional knowledge and functional interrelations, thereby establishing a shared understanding of the system's main properties.
4. Care must be taken that this shared understanding is comprehensible for the different involved stakeholders (e.g. function developers, people in charge of safety, suppliers, etc.) who are often people with a different professional background.
5. Since comprehensibility of the logical architectures is a critical aspect, the logical architecture should also contain non E/E parts, e.g. hydraulics, Bowden cables or environmental objects like the street. The elements shift the focus from modelling only the system under design to a more complete understanding of the system interacting with its environment.
6. Developing a modelling language for logical architectures should be done in line with existing standards and therefore increase the acceptance of the approach.

The demand for a more abstract system description (logical architecture) compared to software and hardware architecture has also been stressed in [15,16]. We follow the author's opinion that an appropriate level of abstraction for modelling logical architectures is the use of logical functions and their exchanged signals disregarding technical details. We denote this kind of logical architecture model "function net". However, logical architectures of complete systems are typically far too complex given the mere number of functions and signals exchanged.

The following problems hinder the usefulness of function nets for complete systems because of the scalability issue.

1. Requirements of *vehicle features* cannot easily be retrieved of the complete function net. Being able to trace vehicle features to logical functions is important because the impact of a requirements change for a feature has to be analysed in the complete function net.
2. The function net contains all functions in all *variants* (sometimes called "150% car"). It does not become clear how a valid configuration of a function net for one car would look like.
3. *Modes* which are internal state changes of functions that result in a major change in the system behaviour (like altered signal communication in case of error degradation) are collectively presented in the static function net.

We propose that vehicle features, variants, and modes can all be captured in a uniform way by using views on the complete function net. Since views can be modelled using the same notation as for complete function nets (as shown in the next section), switching viewpoints between complete function nets and views is possible without difficulty.

The rest of the paper is structured as follows. Section 2 describes how function nets and function net views can be modelled using a variant of SysML Internal Block Diagrams. In the following sections, we show how these views can be used for different purposes: Section 3 introduces views that model vehicle features. In Section 4 views are used to capture variants, and in Section 5 modes are described using views. Section 6 presents related work and Section 7 concludes the paper.

## 2. Complete Function Nets and Views

In order to model logical architectures as function nets, a concrete notation has to be used. There are various options on which a notation for function nets could be based. In [3] we already evaluated UML 2.0 [5] and other UML derivates like UML-RT [13] and SysML [6] which in general were found suitable for architecture and function net modelling [10,11,15]. The detailed reasons for favouring SysML over UML are repeated below:

- SysML uses Systems Engineering terminology. To be more intuitive for people with different professional backgrounds specific terminology of single professional guilds should be avoided. Notations which have their roots in computer science (which use object oriented terminology like classes, objects, associations, etc.) just as well as terminology soley accepted in any other domain, e.g., control theory are likely not to be accepted by different users. SysML already tried to find a good compromise between the different involved terminologies by using a commonly accepted core only.
- SysML requires no strict two layered modelling like in the UML where each structured class consists of parts with no internal structure. This results in more compact diagrams because layers are not distributed among multiple diagrams but can be represented in a single model.
- SysML block diagrams allow us to show communication across multiple hierarchy layers without the explicit use of port delegation. This

simplifies the notation, enhances the readability and avoids the use of unnecessary intermediate signals.

- A SysML block abstracts from the strict instance/type division of the UML which complicates modelling architectures effectively where many elements occur only once in the hierarchy.
- SysML distinguishes between the form of a diagram and its use. This was extremely helpful when we wanted to use the same diagram type with a different semantics. This makes it possible to use one diagram type to model views, variants, modes and complete function nets (as later shown) which improves the compactness of the notation and decreases the necessary training effort.

2.1 Complete Function Nets

Figure 2 shows an example of a function net diagram. The diagram shows a simplified complete function net "CarComfort" which in our case just consists of a model of the central locking functionality. Please note that for organizational reasons diagrams can be split such that many diagrams exist that describe the whole system in diagrams of readable size. The central locking function evaluates the driver's request and closes or opens the doors accordingly. Additionally, an auto lock functionality is modelled, i.e. the doors close automatically if the vehicle exceeds a certain speed limit.

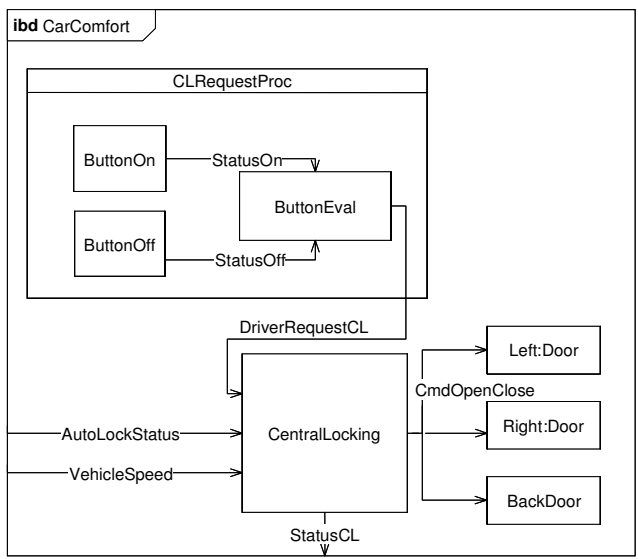

**Figure 2** Function Net Example

Syntactically, the function net is a valid SysML Internal Block Diagram (ibd). In that example, we used three layers of hierarchy (top-level, block "CLRequestProc" and, e.g., "ButtonOn"). With the signal "DriverRequestCL", it also shows an example of cross-hierarchy communication.

For keeping the function net redundant-free and for enabling the re-use of a block multiple times, instantiation is also possible. In the example, there are two doors, which are instantiated by giving each block a name, in this case "left" and "right". These two blocks share their behaviour but not their internal state. A more detailed description about the instantiation mechanism can be found in [3].

We tailored the full SysML Internal Block Diagrams to our specific needs to enable compact definitions and decrease the learning effort for the notation. An Internal Block Diagram that is used as a function net may only contain directed connectors to indicate the signal flow direction. Multiplicities of blocks and connectors are not used in the sense that there is always exactly one block or signal.

2.2 Views of Function Nets

SysML allows us to use the same diagram type for complete function nets and views. Hence views are modelled as Internal Block Diagrams with a few specific properties. Views are always related to some complete function net. Compared to the complete function net, a view which is marked with the stereotype "view" may leave out blocks, signals or hierarchy information. In views, blocks from other features can be imported and marked with a stereotype "ext" to clarify that this block is part of the context of the view and not a central aspect. Additionally, "environmental" blocks and non-signal communication can be added to further increase understandability. The environmental elements refer to non E/E elements that have a physical counterpart.

The example in Figure 3 shows that whole blocks from the complete function net have been left out (block "CLRequestProc") and that the blocks "CentralSettingsUnit" and "VehicleState" have been included in the view to clarify where the signals "AutoLockStatus" and "VehicleSpeed" originate. Also, some environmental elements are included. The physical door look "LockActuator" is shown in the figure and marked with the corresponding stereotype "env". Non-signal communication can be marked with a stereotype "M" for mechanical influence (as shown in the example), "H" for hydraulics, and "E" for electrical interactions.

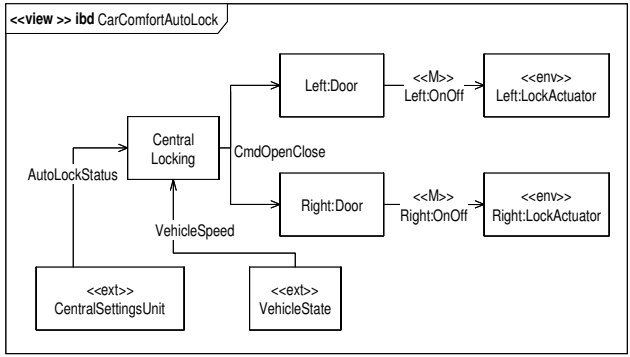

**Figure 3** View on a Function Net

To assure the consistency between a view and the complete function net of the system the following consistency conditions must hold.

1. Each block in the view without a stereotype "ext" or "env" must be part of the complete function net.
2. Whole-Part-Relationships in the view must be present in the complete function net. It is permitted to leave out intermediate layers.
3. Elements that are related via a (possible transitive) Whole-Part-Relationships in the complete function net must also have this relation in a view if both elements are shown.
4. Communication relationships not marked with a stereotype shown in the view must be present in the logical architecture. If the concrete signal is omitted in the view, an arbitrary signal communication must exist in the complete function net.
5. Communication relationships need not be drawn to the exact source or target, any super-block is sufficient if the exact source or target block is omitted.

### 3. Using Views to Describe Features

A feature of a system describes a certain functionality that can be observed by customers. In automotive systems the concrete configuration is usually chosen by the customers when ordering a car. The structure and the interdependencies of features are usually described by feature trees [4] or related notations which help to distinguish valid and invalid combinations of features from each other.

The features of automotive systems can be modelled by interacting functions in logical function nets. The mapping between features and their realising functions may be complex: A feature is usually implemented by a set of interacting functions whereas a single function may play a certain role in the realisation of different features. The mapping is often neglected during later development steps where the developers concentrate on the realisation of the complete function net and discard their initial models. This situation can be avoided by using views which are kept consistent to the complete function net and are therefore a valid and up-to-date model of a single feature throughout the whole development lifecycle.

A new type series is usually planned on a modified feature set of the predecessor by deciding which features remain unchanged and which are omitted or replaced by a refined version. This modified feature set is usually the origin to model the new function net. The re-use of specifications is an advantageous approach because this simplifies the re-use on the subsequent abstraction layers.

Views enable the independent modelling of a feature by showing only the functions and signals that interact to realise a specific feature. Using this idea a set of views specifies the behaviour of a feature in a self-contained description. In addition these views are consistent to a complete function net which models the automotive system.

Another advantage of this approach is that these feature views are a way to model a feature without explicitly showing its realisation within a certain automotive system. Especially the interface of functions is reduced to the necessary parts. These models then can be reused within the next type series which reduces the development cost and increases the quality as already approved descriptions are reused. In addition views allow the developers to analyse the impact of a change request by modifying the view and a subsequent check for consistency with the complete function net. Figure 4 shows the embedding of views in the development process.

In feature views not only software functions are modelled but also physical devices and elements of the environment can be included. This allows the modelling of closed world controllers and increases the understandability of the functionality provided by the feature. The notation also helps the modeller to focus on the central aspects of a feature by using the stereotype "ext". Blocks that are not central to the feature but provide necessary signals are marked as external to indicate that their realisation is not detailed here.

An example for the use of views can be found in Figure 3 where the Auto lock feature with its most relevant functions is explained. Please note that the block VehicleState is marked as external to increase the re-use of this diagram, as the determination of the vehicle speed may vary from type series which is not central to the realisation of this feature as long as there is any subsystem which provides such a signal. The internal behaviour of the block "CentralLocking" would be explained in more detail by another diagram or another suitable textual or model-based specification.

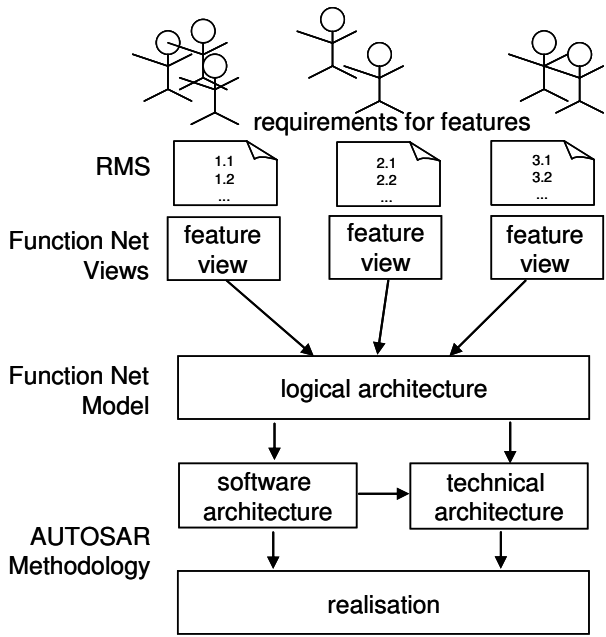

**Figure 4** Example Development Process including views for individual features

## 4. Using Views to Model Variants

Besides describing different features with views independently of the concrete automotive function net in which they are realised, different model variants can be described by a view.

The variants of a feature usually provide the same principle functionality but are distinguishable by having different individual properties. The complete function net includes all variants (the "150% car") and through parameterization the intended variants are chosen. This approach leads to complex function nets that make it difficult to understand which parts are responsible for realising the considered variant when the complete function net is solely used. The individual views for certain variants simplify the understanding because they show a single variant only and are therefore much clearer.

Using this approach, the blocks forming the behaviour of a variant are still localised in the complete function net in the sense that the views for variants are only an extract of the complete function net. An alternative way would be annotating (in form of stereotypes) the variant information to the complete function net and using this function net only. Each block can then be part of some or all variants. This approach turned out to overcrowd the function nets whereas using views facilitates clearer and simpler models.

Examples for using variants are the two versions of the "CentralLocking" subsystem. The basic system does not analyse the vehicle speed and therefore the doors are not closed automatically, whereas the premium function interacts with the other function to close the doors automatically if a certain speed is exceed. Figure 5 shows the two views for the variants of the CentralLocking block by drawing a single diagram for each variant. In the basic view the block "EvalSpeed" is omitted to illustrate that this blocks is disabled in this variant. The incoming signal "VehicleSpeed" is omitted to indicate that it will be ignored by the other blocks and does not play a role in the modelling of this variant. The "OpenClose" signal is omitted to indicate that it will never be sent by this variant.

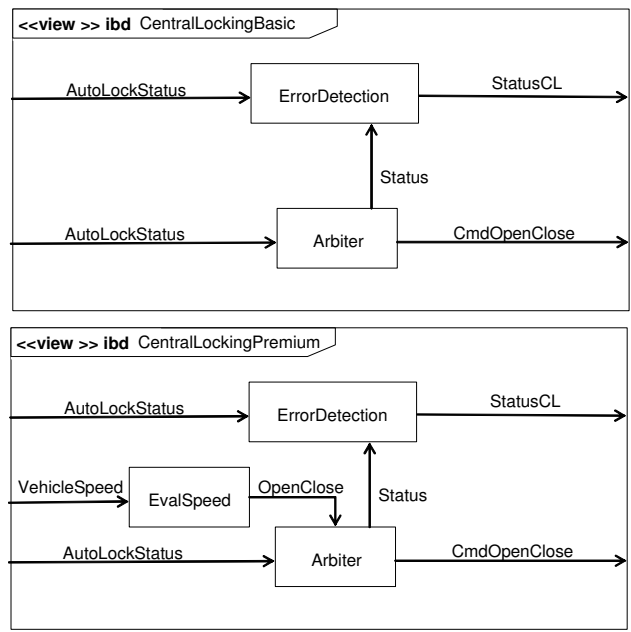

**Figure 5** Using views to model variants

We recommend to describe each variant by a single view that is consistent to the automotive function net by fulfilling the conditions listed in Section 2.2. The view may not be limited to the blocks under design but include the environment of the block to describe the interaction of this block in more detail and allow a better understanding of the variant properties.

## 5. Using Views to Model Modes

Automotive systems are safety-critical systems in the sense that a high reliability has to be assured. Therefore, faults in subsystems usually have to be recognised and handled in order to control the fault's impact on the overall system. In the presence of faults, affected functions should as long as possible provide at least a basic functionality. A complete failure has to be avoided at all costs. In contrast to IT systems where an administrator can be informed about a problem who can take immediate actions if necessary, an automotive system has to work autonomously. Usually as long as possible a safe degradation mode is used where a limited functionality is available until a hardware defect is repaired in the garage.

Modes are an important concept to simplify the description of error degradation in function nets. When a function block changes its operating mode, it usually shows a different behaviour. The different modes of operation can be described by a SysML Statechart where the states are named as the modes of operation. Each of the modes shows a distinct observable behaviour and can be described by an Internal Block Diagram or a view. The transitions are used to describe the conditions which force a change of the mode of operation.

An example for such a Statechart is shown in Figure 6 where two modes are used: One normal mode ("CarComfort") and error degradation ("CarComfortDegradation") where the evaluation of the button results in error values. The transitions between the two states indicate that the degradation mode is entered when one of the input signals "StatusOn" or "StatusOff" shows faulty values. The normal mode is entered again, if the system is reset, e.g. switching the engine of the car off.

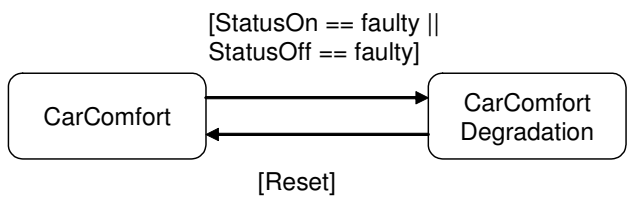

**Figure 6** Error degradation

The normal mode is already described in Figure 2 whereas the degradation is shown in a view that can be found in Figure 7. The block "CLRequestProc" is left out to illustrate that the function is not available in the degradation mode. All incoming signals are removed to indicate that they are ignored. The signal "DriverRequestCL" is omitted because it is not sent in this mode.

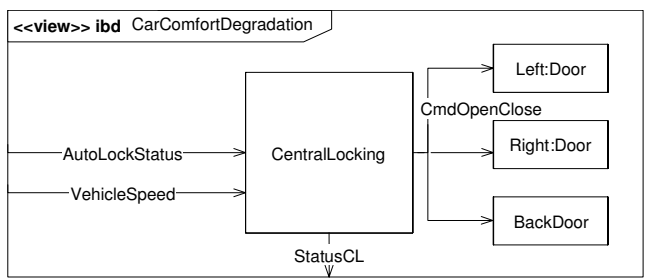

**Figure 7** Using a view to describe a mode

Moreover, the use of modes is not limited to explain the error degradation in automotive systems but it can also be used to describe other modes of operation. A further example would be the different behaviour of the "CentralLocking" block for rear doors if the parental control is activated or not.

## 6. Related Work

In [15] function net modelling with the UML-RT is described. We extended this approach by using the SysML for modelling function nets and explained its advantages. We supplement the approach by views that simplify the transition from requirements to early design phases (feature views), that can capture variants, and modes.

In [12] view merging in the presence of incompleteness and inconsistency is described. The merging algorithm also simplifies the transition from requirements to early design phases like our approach. Especially the constant evolution of requirements during the development makes it difficult to apply such algorithms to our problem. That is why we explicitly decided against automatic merging of views to obtain a complete function net. By checking the consistency between diagrams, consistency problems are found that need to be resolved manually.

In [9,17] service oriented modelling of automotive systems is explained. The service layer is similar to the modelling of features. In addition we explored how services can benefit from modelling the environment together with the feature.

In [18] an alternative approach to the modelling of automotive systems using a formal approach is explained. Modes are defined here without the use of views but with a modularization approach.

In [2] the use of rich components is explained that employ a complex interface description including non-functional characteristics. In contrast to our approach rich components focus less on the seamless transition from requirements to function nets but assume an established predefined partitioning in components.

The AUTOSAR consortium [1] standardises the software architecture of automotive systems and allows the development of interchangeable software components. One main drawback of this approach is that software architectures are too detailed in early development phases where function nets are commonly accepted by developers.

In [7] a notation is introduced how variants can be annotated inside normal software models. The available model elements are marked by a graphical notation and included in the derived variants. The notation is not formalised further and it remains unclear how the approach scales if different variants are not very similar.

## 7. Conclusion

In this paper, we proposed to model the logical architecture of an automotive system as a function net using a SysML-based notation. To increase the usability of the function net we defined views on that function net to alleviate the scalability problem that turns up because of the large number of functions and signals in a complete model. We showed how views can be used to describe vehicle features, variants and modes. Using the same notation for all of these purposes allows for an easy switching of viewpoints.

We concentrated on structural issues and only touched dynamic aspects when we introduced modes. In order to obtain a complete architecture description, also behavioural models may be integrated. It is, for example, possible to add timing or sequence diagrams to feature views to describe typical use cases or scenarios. Annotation of (non-functional) properties (physical, timing, etc.) to function nets can be made to document fixed design decisions or additional constraints. These are directions for further research.